\documentclass[a4paper]{jpconf}
\usepackage{graphicx}
\usepackage{amsmath}
\usepackage{amssymb}

\usepackage{hyperref}

\newcommand{\nua}[1]{\ensuremath{\rlap{\kern-2.5pt\ensuremath{\overset{\scriptscriptstyle(-)}{\phantom{\nu}}}}{\ensuremath{{\nu}_{#1}}}}}
\newcommand{\dm}[1]{\ensuremath{\Delta m_{#1}^2}}
\newcommand{\Us}[1]{\ensuremath{U_{#1}}}
\newcommand{\Usq}[1]{\ensuremath{|U_{#1}|^2}}
\newcommand{\Neff}{\ensuremath{N_{\rm eff}}}

\begin{document}
\title{Light Sterile Neutrinos}

\author{Stefano Gariazzo}

\address{Istituto Nazionale di Fisica Nucleare (INFN), Sezione di Torino, Via P. Giuria 1, I-10125 Turin, Italy}
\ead{gariazzo@to.infn.it}

\begin{abstract}
We review the status of light sterile neutrino searches,
motivated by the original Short BaseLine (SBL) anomalies.
Here, we discuss how sterile neutrino properties can be constrained by different types of neutrino oscillation experiments (considering appearance or disappearance probes in different oscillation channels) and non-oscillation measurements.
These latter include experiments aiming at obtaining a value for the absolute scale of neutrino masses ($\beta$ decay probes)
and the indirect constraints that we could obtain from cosmological observations.
\end{abstract}

\section{Anomalies}
\label{sec:anomalies}
Neutrino oscillations certify us that neutrinos are massive particles,
that can be described according to the way
they interact (flavor eigenstates) or propagate (massive eigenstates).
These two descriptions are related by a mixing matrix.
Considering three neutrinos,
our kwnoledge of neutrino oscillations is nowadays quite good, see e.g.\cite{deSalas:2020pgw}:
we have reached a percent precision on several mixing parameters
and most experiments are in agreement with the theoretical predictions.
On the other hand, there are still some unknowns, such as
the value of the Dirac CP violating phase,
the mass ordering (i.e.\ the sign of \dm{31}~%
\footnote{We define $\dm{ij}=m^2_i-m^2_j$.}
)
and the octant of the $\theta_{23}$ angle.

Despite the level of precision,
not all the experimental results are successfully described in the context of a three-neutrino paradigm.
In particular, four anomalies have been thoroughly debated and investigated over the last twenty years (see e.g.\ \cite{Gariazzo:2015rra,Giunti:2019aiy}):
\begin{itemize}
\item the LSND experiment \cite{LSND:2001aii} observed an excess appearance of $\bar\nu_e$ events in a beam of $\bar\nu_\mu$, with a significance of $\approx3.8\sigma$;
\item an anomalous disappearance of electron neutrinos was observed during the calibration of the Gallium solar experiments GALLEX and SAGE \cite{Giunti:2010zu}, with a significance of almost $3\sigma$;
\item after the new evaluations of the reactor antineutrino fluxes by Huber \cite{Huber:2011wv} and the Mueller group \cite{Mueller:2011nm},
a discrepancy between the predicted electron antineutrino rate and the observations at several reactor experiments emerged, with a significance of $\approx3\sigma$ \cite{Mention:2011rk};
\item the MiniBooNE experiment \cite{MiniBooNE:2020pnu}, originally built to check the LSND anomaly,
finally reported an excess appearance of events with a significance of almost $5\sigma$.
The excess is mainly present at the lowest energies probed by MiniBooNE, i.e.\ below 400~MeV.
\end{itemize}
There is a common aspect in all these experiments: they observe neutrino oscillations at Short BaseLines (SBL),
where ``baseline'' indicates the ratio between travelled distance and neutrino energy ($L/E$).
None of these experimental results can be explained in the context of three-neutrino oscillations,
which cannot develop at the $L/E$ tested by these probes,
and researchers have been trying to find a solution to these anomalies for many years.

When considering the standard model of particle physics,
we know that the number of neutrinos which have interactions with the $Z$ boson,
also named ``active neutrinos'', is three~\cite{ALEPH:2005ab,Janot:2019oyi},
but the theory actually allows the presence of an unlimited number of ``sterile'' species,
intended as right-handed singlets, which cannot have weak interactions.
These sterile neutrinos would however oscillate with active neutrinos,
and in such way they can have an impact on the observables we test at experiments.

For several years, a model with three active and one sterile neutrino,
separated by a mass splitting of approximately
$\dm{\rm SBL}\sim1$~eV$^2$,
has been considered
a possible solution to the abovementioned SBL anomalies,
although, as we will see, nowadays it is difficult to explain
all the currently available experimental results within this scenario.
In any case, here we will consider the so-called ``3+1'' scenario,
where the fourth neutrino is mostly mixed with the sterile flavor (the ``+1'')
and it is much heavier than the first three mass eigenstates,
mostly mixed with active neutrino flavors and characterized by a mass much smaller than 1~eV.
Despite this, we denote the sterile neutrino as ``light'', since its mass, at the eV scale,
is still much smaller than the mass of the other known fermions and bosons in the Standard Model.
Sterile neutrinos with larger masses, see e.g.\ \cite{Adhikari:2016bei},
are beyond the scope of the present discussion.

Assuming the 3+1 scenario, the $4\times4$ neutrino mixing matrix can be described by a total of
6 mixing angles
(including three new angles $\theta_{14}$, $\theta_{24}$, $\theta_{34}$, which are assumed to be small),
3 Dirac CP violating phases (including two new phases)
and 3 Majorana phases (including one new phase).
For SBL oscillations, we are mostly interested in the last column of the mixing matrix,
which can be written as
$\Usq{\alpha4}
=
[s^2_{14},\,
c^2_{14}s^2_{24},\,
c^2_{14}c^2_{24}s^2_{34},\,
c^2_{14}c^2_{24}c^2_{34}]$~%
\footnote{We define $s_{ij}=\sin\theta_{ij}$ and $c_{ij}=\cos\theta_{ij}$.},
where in this case $\alpha\in[e,\mu,\tau, s]$.
When considering SBL neutrino oscillations,
the transition probability between some flavor $\alpha$ at the source
and a flavor $\beta$ at the detector
can be written as a function of the
$\Usq{\alpha4}$ mixing matrix elements alone, where $\alpha\in[e,\mu,\tau]$.
The effect of the Dirac phases is not visible at SBL and must be studied at Long BaseLine experiments (see e.g.\ \cite{Palazzo:2020tye}).
The assumptions that $m_4\gg m_i$ ($i\in[1,2,3]$) and that the angles $\theta_{14}$, $\theta_{24}$, $\theta_{34}$ are small
are required in order not to spoil the phenomenology of three-neutrino oscillations:
the fact that $\nu_4$ is heavier than the other states implies that the new oscillation channels generate very fast oscillations,
that are only visible at small $L/E$, while they are averaged out at large distances.
The amplitude of the oscillations, proportional to the matrix elements,
is instead required to be small so that the transition probabilities
of three-neutrino oscillations are almost unchanged at large distances,
where the new mixing angles enter mostly through their cosine (which is close to one).

Notice that the latter assumption is not automatically verified.
The MiniBooNE experiment, for example, prefers maximal mixing between active and sterile neutrinos \cite{MiniBooNE:2020pnu}:
the best-fit of the most recent data is obtained
for values of the effective angle $\sin^22\theta_{e\mu}\approx1$.
Since this mixing angle is proportional to the product of \Usq{e4} and \Usq{\mu4},
MiniBooNE measurements are not compatible
with the assumption that \Usq{e4} and \Usq{\mu4} are small.
Moreover, even considering maximal active-sterile mixing, the observations in the low-energy bins
exceed the theoretical prediction, so that a full sterile neutrino solution to MiniBooNE appears unlikely.
The origin of the MiniBooNE low-energy excess
is under investigation at MicroBooNE, see e.g.~\cite{MicroBooNE:2016pwy,MicroBooNE:2021zai}.

\section{Beta decay probes}
\label{sec:beta}
The existence of any neutrino mass eigenstate is in principle visible in the kinematics of processes such as $\beta$ decay.
If one considers an atom of tritium, for example, the $\beta$ decay process takes place when the tritium atom decays into a light Helium, emitting one electron and one antineutrino.
The two leptons carry away most of the available kinematic energy.
Using the fact that the minimum antineutrino energy depends on its mass, one can describe the end-point of the electron energy spectrum using the Kurie function \cite{Giunti:2007ry}:
\begin{equation}
K(T)
=
\left[
(Q_\beta-T)
\sum^N_{i=1}\Usq{ei}
\sqrt{(Q_\beta-T)^2-m_i^2}
\right]^{1/2}\,,
\end{equation}
where $T$ is the electron kinetic energy,
$Q_\beta$ is the $Q$-value of the reaction
and $m_i$, \Usq{ei} are the mass of the $i$-th massive eigenstate and its mixing with the electron neutrino flavor.
Since the shape of the electron spectrum at the endpoint depends on the neutrino masses and mixings of the $N$ neutrinos,
therefore, experiments such as KATRIN can access to the absolute scale of the neutrino masses~\cite{Aker:2021gma},
and also study the presence of additional mass eigenstates.

The KATRIN collaboration published their first constraints on the mass splitting and mixing of one light sterile neutrino with the electron flavor in \cite{KATRIN:2020dpx}, see also \cite{Giunti:2019fcj}.
The results mostly constrain mass splittings \dm{41} between a few eV$^2$ and approximately $10^3$~eV$^2$, because of the observed energy window around the endpoint of the electron spectrum.
Although current constraints are not yet competitive with probes at reactor experiments (see section~\ref{sec:dis_e}), the final KATRIN sensitivity is expected to probe mixing angles a factor 10 smaller than current bounds.
Such level of precision will be sufficient to either confirm or reject the preferred region by Neutrino-4 \cite{Serebrov:2020kmd}.

\section{Cosmology}
\label{sec:cosmo}
Despite the fact that it cannot interact with the standard model particles through weak interactions,
the sterile neutrino can be efficiently produced in the early universe thanks to its oscillations with active neutrinos.
Neutrino oscillations, however, cannot take place while interactions are sufficiently frequent to maintain neutrinos in the flavor eigenstates: this happens when the density of the thermal plasma is very high.
Consequently, in the very early universe the sterile neutrino cannot be produced,
until oscillations generated by \dm{41} are no longer blocked by the matter effects of the thermal plasma.
In particular, neutrino oscillations corresponding to higher mass splittings start to occur earlier.

Having a mass of approximately 1~eV, the fourth neutrino would be relativistic at the time of neutrino decoupling and Big Bang nucleosynthesis, and possibly until matter-radiation equality.
This means that it contributes to the total amount of radiation (relativistic particles) in the early universe.
It is convenient to parameterize the contribution of neutrinos to the total amount of radiation by means of the effective number of relativistic species, \Neff, which accounts for the number of neutrino-equivalent particles.
Due to the fact that neutrino decoupling does not take place instantaneously, active neutrinos contribute with $\Neff=3.044$ \cite{Akita:2020szl,Froustey:2020mcq,Bennett:2020zkv},
slightly larger than three because of the entropy transfer from electron to high-momentum neutrinos.
The sterile neutrino, if in full equilibrium with the active species,
would raise such number to $\sim4$.

The thermalization of the additional neutrino, as already mentioned, depends on the SBL mass splitting and on the new mixing angles (or the mixing matrix elements \Usq{\alpha4}, with $\alpha\in[e,\mu,\tau]$).
The dependence of \Neff\ on the mixing parameters has been studied for several years.
While the first results were obtained using some kind of approximation,
for example considering a simplified 1+1 (one active plus one sterile neutrino) scenario \cite{Dolgov:2003sg,Hannestad:2012ky},
the most recent numerical calculations~\cite{Gariazzo:2019gyi} can fully take into account a 3+1 case and the effect of each mixing angle appropriately.
This is done by means of the numerical code \texttt{FortEPiaNO}%
\footnote{Publicly available at \url{https://bitbucket.org/ahep_cosmo/fortepiano_public}.},
which allows to compute \Neff\ as a function of all the mixing parameters.
Our results show that,
for large mass splittings \dm{41},
\Neff\ is almost independent of the mixing angles between active and sterile neutrino,
and confirm that cosmology is practically insensitive to neutrino flavors.
For a wide fraction of parameter space, the thermalization of the sterile neutrino is complete,
and $\Neff\simeq4$.
Moreover, oscillations are more effective when more than one active-sterile mixing angle is different from zero, since oscillation channels work in parallel and more easily bring the fourth neutrino in equilibrium with the standard ones.

The amount of radiation energy density is very well constrained by observations of the Cosmic Microwave Background (CMB).
The most recent measurements by the Planck experiment exclude $\Neff=4$ at a significant level \cite{Planck:2018vyg}:
this means that the presence of a fully-thermalized sterile neutrino is disfavored.
Using CMB constraints, therefore, it is possible to derive bounds on the mixing matrix elements and mass splitting between active and sterile neutrino:
the most recent cosmological constraints translate into bounds $\Usq{\alpha4}\lesssim10^{-3}$ \cite{Hagstotz:2020ukm}, which is significantly stronger than the bounds obtained by terrestrial experiments, as we will discuss in what follows.

\section{Disappearance, muon channel}
\label{sec:dis_mu}
Disappearance of muon (anti)neutrinos can be probed using atmospheric neutrinos
or neutrino beams produced at accelerators.
Since the involved energies are typically much larger than those achieved at reactors,
the considered distances are also much larger.
Simple SBL oscillations in the $\nua{\mu}$ disappearance channel would mostly probe the \Usq{\mu4} mixing matrix element,
but accelerator and neutrino experiments generally have a limited sensitivity also on \Usq{\tau4}.
Moreover, Long BaseLine (LBL) measurements considering accelerator neutrinos have access
to the two additional Dirac CP phases associated to the \Us{e4} and \Us{\tau4} matrix elements \cite{Palazzo:2020tye},
which however are not discussed here.

One of the most important experiments probing \nua{\mu} disappearance is
MINOS/MINOS+~\cite{MINOS:2016viw,MINOS:2017cae},
detecting neutrinos at distances of $\approx500$~m (near detector) and $\approx800$~km (far detector)
from the beam source, which emits neutrinos at energies of a few GeV.
The most recent MINOS+ observations, obtained by performing a full fit of the data
collected by the two detectors,
put strong bounds on \Usq{\mu4}, which is constrained to be smaller than approximately $10^{-2}$ (90\% CL) over a wide range of mass splittings.

Atmospheric neutrino observations, on the other hand, consider neutrinos crossing the entire atmosphere and the Earth, from all the possible direction:
they have access to a variety of possible distances.
One of the main experiments is IceCube,
consisting in several strings of spherical optical sensors inserted in the ice, in Antarctica.
IceCube puts constraints on active-sterile neutrino mixing
thanks to two types of observations:
high energy events ($\gtrsim300$~GeV) \cite{IceCube:2016rnb}, which are detected using the full size of the experimental apparatus,
plus neutrinos at energies $\mathcal{O}(10\text{ GeV})$ \cite{IceCube:2017ivd},
which instead are tested by the inner and denser section of the detector, called DeepCore.
The results published until 2020 showed no sign of active-sterile neutrino oscillations
and put bounds on the mixing parameters, which in the case of \Usq{\mu4} are comparable with the ones from MINOS/MINOS+ at mass splittings of a few 0.1~eV$^2$.
The most recent results~\cite{IceCube:2020phf}, obtained studying 8~years of neutrino data,
suggested instead the existence of active-sterile neutrino oscillations,
although with a mild significance.
This was the first time a signal in favor of sterile neutrinos has been observed in the muon (anti)neutrino disappearance channel.

Recently, the NO$\nu$A collaboration also published constraints on active-sterile mixing \cite{NOvA:2021smv}.
Although their bounds on \Usq{\mu4} and \Usq{\tau4} are not competitive with those obtained by other experiments,
it is the first time that the a LBL experiment is able to constrain such properties.

\section{Electron (anti)neutrino disappearance}
\label{sec:dis_e}
The electron (anti)neutrino disappearance channel is the one where most of the activity
has been concentrated in the last years.
Two anomalies enter this category: the reactor antineutrino anomaly (RAA) and the Gallium anomaly.

As already mentioned,
the RAA finds its origin in the 2011 calculations of the reactor antineutrino fluxes
by Huber \cite{Huber:2011wv} and the Mueller group \cite{Mueller:2011nm},
denoted ``HM'' model in the following.
The RAA was discovered when comparing existing reactor observations
with the predictions obtained using the HM fluxes:
the original anomaly
indicated a disappearance of electron antineutrinos,
with a significance of about $3\sigma$~\cite{Mention:2011rk}.
Notice that the RAA only takes into account the observation of the rate of antineutrinos
by a number of experiments, most of which could only observe the total rate, but not the shape of the spectrum.

While at first the RAA was explained by assuming that the observed rates were reduced
due to active-sterile neutrino oscillations,
in the following years one important question started to arise:
do we trust the theoretical antineutrino fluxes enough to trust the anomaly?
The doubt was legitimate, in particular after the observations, by different experiments,
of a shape distortion that affects the antineutrino spectrum at energies between approximately
4 and 6~MeV:
this excess of events is known as the ``5~MeV bump'' or ``shoulder'' \cite{DoubleChooz:2014kuw,DayaBay:2015lja,RENO:2015ksa}.
Since its discovery, several studies have been devoted to understand the 5~MeV bump,
but a conclusive explanation has not been found yet.
One of the possibilities is of course that there is something wrong
in the theoretical calculation of the HM fluxes.
If that is the case, we should not trust the RAA
at the time of constraining sterile neutrino properties.

The theoretical study of the reactor antineutrino flux has been revisited by several authors after the publication of the HM fluxes.
Such studies are normally performed in two ways.
The method adopted by \cite{Huber:2011wv} is called ``conversion method'',
and it is based on measurement of the electron energy spectrum in order to reconstruct the neutrino one.
The conversion method takes into account the sum of a number of virtual $\beta$-decay branches,
described by parameters which can be obtained by fitting the available electron data.
Each virtual $\beta$ branch is later converted into the corresponding neutrino branch according to nuclear theory.
This method needs as an input the measurement of the total $\beta$ spectrum from reactors,
which in the case of Ref.~\cite{Huber:2011wv}
was the spectrum measured at ILL in the 1980s \cite{Hahn:1989zr,Schreckenbach:1985ep,VonFeilitzsch:1982jw}.
On the other hand, \cite{Mueller:2011nm} used a method called ``ab-initio'',
which aims at reconstructing the total reactor energy spectrum by summing up
the energy spectra of
all the known $\beta$-decays of Uranium/Plutonium and of their decay products,
weighed according to the corresponding yields.
The very recent paper \cite{Giunti:2021kab} analysed how the significance of the RAA
changes when different theoretical fluxes are considered.
By taking into account all the available rate measurements from reactor experiments,
the authors quantify the RAA computed using the HM fluxes to have a
significance of $\sim2.5\sigma$.
When considering the spectrum obtained by \cite{Hayen:2019eop},
which uses a conversion method based on the same ILL data also employed by \cite{Huber:2011wv},
but allowing for forbidden transitions,
the RAA reaches a larger significance of 2.9$\sigma$.
Using the ab-initio method,
the authors of \cite{Estienne:2019ujo} obtained a reactor antineutrino spectrum
which is denoted as EF, for which the significance of the anomaly decreases to 1.2$\sigma$,
meaning that the expectations are almost in agreement with observations.
The agreement between theory and expectation is improved even more
when one considers the very recent calculation of the reactor spectra by~\cite{Kopeikin:2021ugh},
based on the conversion method applied on new Kurchatov Institute (KI) measurements
\cite{Kopeikin:2021rnb}
of the electron spectrum.
Due to the discrepancy between the new KI measurements and previous ones at ILL,
indeed, the anomaly significance is reduced to 1.1$\sigma$.
If these last evaluations of the RAA are correct, there is no need to advocate
the presence of a sterile neutrino in order to reconcile theory and observations.

Due to the lack of confidence on the absolute reactor antineutrino flux,
one of the best ways to study the presence of neutrino oscillations at SBL reactor experiments
is to take measurements at different distances, and then consider the ratio between the various observed fluxes.
In such way, the absolute neutrino flux can be factorized out and
the results are model-independent,
in the sense that oscillation constraints are not affected by
the theoretical calculations of the reactor antineutrino flux.
Over the last few years, several experiments explored this direction,
working in different ways:
using multiple detectors
(for example NEOS normalizes its observations using results from DayaBay~\cite{NEOS:2016wee}
or RENO~\cite{Atif:2020glb})
using segmented detectors
(PROSPECT~\cite{PROSPECT:2020sxr},
STEREO~\cite{STEREO:2019ztb})
or a single detector which can be moved at different positions
(DANSS \cite{DANSS:2018fnn} and
Neutrino-4~\cite{Serebrov:2020kmd}, see also \cite{Almazan:2020drb,Serebrov:2020yvp,Giunti:2021iti}).
Over the years, several among these experiments claimed a preference in favor of active-sterile oscillations
over the three-neutrino scheme,
although the various experiments are not always in agreement with one another:
the best-fit values obtained by Neutrino-4~\cite{Serebrov:2020kmd},
for example, are incompatible with the bounds obtained by PROSPECT~\cite{PROSPECT:2020sxr}.
The statistical significance of the model-independent results
from reactor experiments at SBL has also been strongly debated.
A recent paper \cite{Giunti:2020uhv} demonstrated that even if the true number of neutrinos
is three, it is possible that the best-fit to the observations
indicates active-sterile mixing parameters different from zero,
because the statistical fluctuations of noise can be better fitted
by an oscillating pattern rather than by a constant.
For this reason, the statistical significance is overestimated
if the limits on the oscillation parameters are computed using the Wilk's theorem
instead of the true $\chi^2$ distribution.
Since this must be obtained by simulating toy experiments several times,
the process can be computationally expensive.
When the $\chi^2$ is properly taken into account,
from a combination of all the SBL results that are not in tension with one another
it emerges that the analysis
prefers active-sterile oscillations 
versus the three-neutrino case,
with a best-fit at $\dm{41}\simeq1.3$~eV$^2$, $\Usq{e4}\simeq0.007$
and significance of something less than $2\sigma$ \cite{Giunti:2020uhv}.

The second anomaly in the electron (anti)neutrino disappearance channel
is the Gallium anomaly, from the calibration of the GALLEX and SAGE experiments
\cite{Abdurashitov:2005tb,Giunti:2010zu}.
The original significance of the anomaly, based on cross-section calculations from \cite{Krofcheck:1985fg,Frekers:2011zz}, was approximately~$2.9\sigma$.
When revisiting the experimental constraints using more recent cross section estimates,
the significance decreases to approximately $2.3\sigma$~\cite{Kostensalo:2019vmv}.
In any case, recent evaluations of the Gallium anomaly,
including data from the BEST experiment \cite{Barinov:2021asz},
are in tension with the RAA constraints on the active-sterile mixing matrix element \Usq{e4} \cite{Giunti:2021kab}.

\section{Global fits}
\label{sec:global}
In order to understand the viability of the light sterile neutrino
as a possible solution of the SBL anomalies,
we must perform combined analyses where we consider all the available experimental constraints
at the same time.
Among the last ``global analyses'', we find \cite{Gariazzo:2017fdh,Dentler:2018sju},
which however are a few years old.
The reason for which there are no recent global fits is related to what we will discuss in the rest of this section: the so-called appearance-disappearance tension.

Before attempting a global analysis, let us consider separately the classes of experiments
that observe the same channel, starting from the appearance case.
As we have already mentioned, two of the SBL anomalies come from appearance probes:
by the LSND \cite{LSND:2001aii} and MiniBooNE \cite{MiniBooNE:2020pnu} experiments.
These two are in partial agreement on the preferred mixing parameter space,
but part of the best-fit region is actually in tension with other observations.
These include ICARUS \cite{ICARUS:2013cwr} and OPERA \cite{OPERA:2018ksq},
which exclude large active-sterile mixing angles
by not observing anomalous appearance of electron neutrinos in a muon neutrino beam.
As a consequence, the combination of appearance probes indicates a favored region
at effective mixing
angles $10^{-3}\lesssim\sin^22\theta_{e\mu}\lesssim10^{-2}$
and mass splittings $0.3\lesssim\dm{41}/\text{eV}^2\lesssim1.5$
at $3\sigma$.
Recall that the effective mixing angle $\sin^22\theta_{e\mu}$ is proportional to
the product of \Usq{e4} and \Usq{\mu4}.

Concerning the muon (anti)neutrino disappearance channel,
the constraints from all the available observations agree in setting a rather strong bound on \Usq{\mu4},
which must be smaller than approximately $10^{-2}$ at $3\sigma$
over a wide range of mass splittings.
The constraints are dominated by MINOS/MINOS+ \cite{MINOS:2016viw,MINOS:2017cae}
in most of the considered range for \dm{41}.
The analysis does not take into account the most recent results from IceCube \cite{IceCube:2020phf},
which were released to the general public only a few days before the publication of this proceeding,
in the form of a two-dimensional $\chi^2$ table as a function of \dm{41} and \Usq{\mu4}.
Although not important for our discussion of the appearance-disappearance tension,
muon (anti)neutrino disappearance probes also
set constraints on the \Usq{\tau4} matrix element.

The last set of oscillation results includes reactor and Gallium probes of electron (anti)neutrino disappearance.
In this case, as we discussed extensively in section~\ref{sec:dis_e},
the situation is nowadays rather unclear.
We have some probes claiming a strong preference in favor of a large \Usq{e4}
(such as Neutrino-4 \cite{Serebrov:2020kmd} and BEST \cite{Barinov:2021asz}),
while other probes exclude these results
(mostly reactor experiments such as PROSPECT \cite{PROSPECT:2020sxr}
and the recent evaluations of the RAA from \cite{Giunti:2021kab}).
Combining all the reactor constraints that are not in tension with one another,
the preferred best-fit point \cite{Giunti:2020uhv} has
$\dm{41}\simeq1.3$~eV$^2$ and $\Usq{e4}\simeq0.007$
(the upper limit at $3\sigma$ is around 0.015 at the best-fit value of \dm{41}).

Once we combine these three series of results,
we easily see that the constraints on \Usq{e4} and \Usq{\mu4} cannot be simultaneously satisfied:
if we consider the upper limits obtained by disappearance probes,
we have a very small $\sin^22\theta_{e\mu}\lesssim7\times10^{-4}$.
Such bounds are
in tension with appearance constraints which prefer a large mixing,
in particular LSND \cite{LSND:2001aii} and MiniBooNE \cite{MiniBooNE:2020pnu}.
This incompatibility has been called the ``appearance-disappearance tension'',
and has nowadays a statistical significance that goes well beyond ``strong''
\cite{Dentler:2018sju,Diaz:2019fwt}.

When going beyond oscillation data, there is also a problem with cosmological data.
As we have discussed in section~\ref{sec:cosmo},
CMB observations prefer an incomplete thermalization for the sterile neutrino.
On the other hand, the theoretical predictions for \Neff\ point towards a fully thermalized sterile state,
regardless they are computed considering the mixing parameters that emerge from
the analysis of reactor data,
of appearance constraints
or the recent hints from IceCube.
In other words, given a standard thermalization procedure due to neutrino oscillations,
the cosmological bounds prefer mixing matrix elements
$\Usq{\alpha4}\lesssim10^{-3}$ ($\alpha\in[e,\mu,\tau]$) \cite{Hagstotz:2020ukm},
much smaller than any of the hints we obtain from oscillation experiments.

\section{Conclusions}
\label{sec:conclusions}
Given our current knowledge of neutrino oscillations and cosmology,
the situation is not favorable for the light sterile neutrino.
Non-oscillation probes such as kinematics of tritium $\beta$ decay and cosmology
only put strong constraints on the existence of a new neutrino species.
Many hints in favor of its existence were proposed by a number of different experiments.
Some of these hints are more significant than others,
reaching a significance of nearly $5\sigma$ in some cases,
but few of them are in agreement with one another.
We observe these tensions
between different experiments
(Neutrino-4 and PROSPECT,
LSND/MiniBooNE and ICARUS/OPERA)
or
different classes of probes
(the reactor anomaly versus the Gallium anomaly,
the appearance-disappearance tension,
the incompatibility between cosmological and oscillation constraints).

Whatever we are observing,
we most likely need some kind of new physics to explain all the results.
Such new physics may be something we still miss in our theoretical models
or some problem with the interpretation of observations.
In any case, as of today,
it appears that such new physics may be something else
than a sterile neutrino with a mass of about 1~eV,
and we need future experiments to guide us towards a deeper understanding
of theory and observations.

\ack
SG acknowledges financial support from the European Union's Horizon 2020 research and innovation programme under the Marie Skłodowska-Curie grant agreement No 754496 (project FELLINI).

\section*{References}
\bibliography{main}

\end{document}